\documentclass[aps,pre,superscriptaddress,showkeys,reprint]{revtex4-2}
\usepackage{amsmath,amssymb}
\usepackage{graphicx}
\usepackage{dcolumn}
\usepackage{bm}
\usepackage[latin1]{inputenc}
\usepackage{float}
\usepackage{color}
\usepackage{booktabs}
\usepackage{amsmath}
\usepackage[toc,page]{appendix} 
\usepackage{latexsym}
\usepackage{amsmath}
\usepackage{mathrsfs}
\usepackage{bm}
\usepackage{amsthm}
\usepackage{physics}
\usepackage{adjustbox}
\usepackage{bbm}
\usepackage{amsfonts}
\usepackage{amssymb}
\usepackage{color}
\usepackage{epsfig}
\usepackage{hyperref}
\hypersetup{colorlinks=true, citecolor=blue}
\usepackage{soul}
\usepackage{cancel}
\usepackage[normalem]{ulem}
\usepackage[cmtip,all]{xy} 
\usepackage{comment}
\usepackage{multirow}
\usepackage{mathtools}
\usepackage{dsfont}

\begin{document}

\title{Minimal Model for Chirally Induced Spin Selectivity: Spin-orbit coupling, tunneling and Decoherence}

\author{Miguel Mena}
\affiliation{Departamento de F\'isica, Colegio de Ciencias e Ingenier\'ia, Universidad San Francisco de Quito, Diego de Robles y V\'ia Interoce\'anica, Quito, 170901, Ecuador}

\author{Bertrand Berche}
\affiliation{Laboratoire de Physique et Chimie Theorique, Universite de Lorraine, CNRS, Nancy, France}

\author{Ernesto Medina}
\affiliation{Departamento de F\'isica, Colegio de Ciencias e Ingenier\'ia, Universidad San Francisco de Quito, Diego de Robles y V\'ia Interoce\'anica, Quito, 170901, Ecuador}

\date{\today}

\begin{abstract}
Chirally Induced Spin Selectivity (CISS) is a transport phenomenon observed in both linear and non-linear regimes, where the spin-orbit coupling (SOC) acts as the key driver and electron tunneling serves as the dominant mechanism for charge transfer. Despite SOC's inherent time-reversal symmetry (TRS) preservation, conventional reciprocity relations limit spin polarization and the differential treatment of spin species. In experimental systems, an additional factor, spin-independent decoherence, disrupts TRS and reciprocity. We introduce decoherence using the Buttiker voltage probe within the scattering matrix framework. Our results reveal the importance of under-the-barrier decoherence as an order-of-magnitude polarization enhancement mechanism. Polarization arises by the disruption of spin precession around the spin-orbit magnetic field with a new spin component along the field direction. The alignment of polarization depends on interference effects produced by the voltage probe. We discuss the connection of our model to a more realistic decoherence mechanism in molecular systems. 
\end{abstract}

\date{June 2021}

\maketitle

\section{Introduction}

Chirally Induced Spin Selectivity (CISS) has attracted considerable interest in the literature since it shows spin accumulation induced by chiral molecules in a two-terminal setup\cite{NaamanScience, NaamanDNA,HeliceneKiran,HeliceneZacharias}. Generally, the chiral molecules involved are organic, with no exchange interactions present and no magnetic materials involved, so the spin-orbit (SO) coupling is assumed to be the only spin-active\cite{Yeganeh2009} interaction, albeit weak. This apparently contradicted the large spin polarization effects observed, stronger than those of ferromagnets.
On the other hand, the two-terminal, quasi-one-dimensional setup, in the presence of time-reversal symmetric couplings (such as SO), should satisfy reciprocity relations\cite{Kiselev,Bardarson2008,Bart1}, forbidding two terminal polarization. So, all numerical models yielded too small a polarization and mysteriously violated reciprocity. Interface effect, typically with metals, promised an explanation for the SO size. Nevertheless, the effect has been shown to be present even for weak SO metal contacts\cite{HeliceneZacharias}. 

Here we show that adding two physically motivated ingredients to the model accounts for the previous issues and inspires model detailed models for more quantitative estimates: i) Tunneling is a well-known electron transfer in molecules and should be accounted for in the model\cite{Nitzan2001}. ii) Most experiments are at room temperature so coupling to the environment or additional reservoirs are expected beyond the two terminals. Such couplings would imply a system beyond two terminal reciprocity that breaks time-reversal symmetry.

This work will discuss the simplest transmission model through an SO active barrier. We will explore the possibilities of spin-polarized electron transmission in a one-dimensional two-probe setting for an exactly solved model. The action of a $U(1)$ field on tunneling electrons\cite{Buttiker} under a barrier relaxes the electron spin toward the lower energy orientation. This behavior will be contrasted with the effective momentum-dependent magnetic field arising from the SO coupling. A critical issue in the latter case is the velocity operator's non-diagonal nature that secures the flux's continuity through boundaries\cite{Molenkamp}. The strong result is that while a $U(1)$ magnetic field (breaking TRS) polarizes spin along its direction, under the action of the barrier, no spin polarization results under the action of a spin-orbit ``magnetic field" as expected from reciprocity. Breaking TRS weakly, we show that spin polarization arises over and under a barrier. Still, under the barrier, the effect is enhanced by orders of magnitude, fitting the physical picture of transport in molecules. Our implementation of Buttiker's probe with spin will break TRS by dephasing under the barrier. while bove the barrier, there is also a substantial non-hermiticity in the Hamiltonian by leaking probability to the environment. These findings address core issues in our understanding of the CISS effect.







\section{Barrier Model with a Rashba term: Full Coherence}
\begin{figure}[ht]
\includegraphics[width=8.6cm]{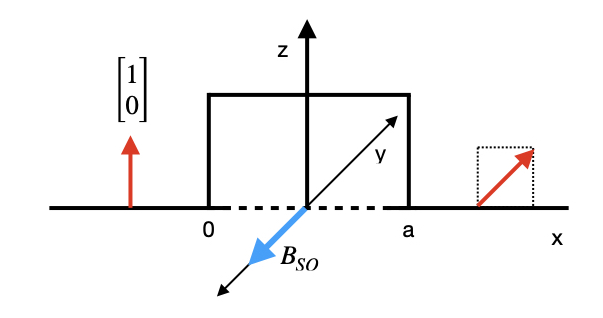}
    \caption{Spin tunneling setup chosen: We inject a spin-up electron in the $z$-axis into a spin active barrier with a $B_{SO}$ in the $-{\hat y}$ direction.}
    \label{SpinTunnelingSetup}
\end{figure}

Following reference \cite{MedinaBerche}, we solve the one-dimensional scattering problem in Fig.\ref{SpinTunnelingSetup} for the model
\begin{equation}
    \cal{H}=\begin{cases}
			(\frac{p_x^2}{2m}+V_0)\mathds{1}_{\sigma}+\Lambda p_x\sigma_y, & \text{if~ $0<x<a$}\\
            (\frac{p_x^2}{2m})\mathds{1}_{\sigma}, & \text{otherwise},
		 \end{cases}
		 \label{Hamiltonian2}
\end{equation}
where $\mathds{1}_{\sigma}$ is the unit matrix in spin space and $\sigma_i$ are the Pauli spin matrices. $\cal{H}$ acts on the spinors $\psi=\left(\psi_+(x)~\psi_-(x)\right)$ where $|\psi_{\pm}|^2dx$ is the probability of find a particle between $x$ and $x+dx$ with spin projection $\pm \hbar/2$. This Hamiltonian contains only the spin active part of the helical model, so no orbital angular momentum is present\cite{VarelaZambrano,Varela2016,Oligopeptides2020}. Nevertheless, the magnitude of the SO coupling is derived from the overlaps pertaining to the chiral structure considered in those models.
\begin{figure}
\includegraphics[width=8.0cm]{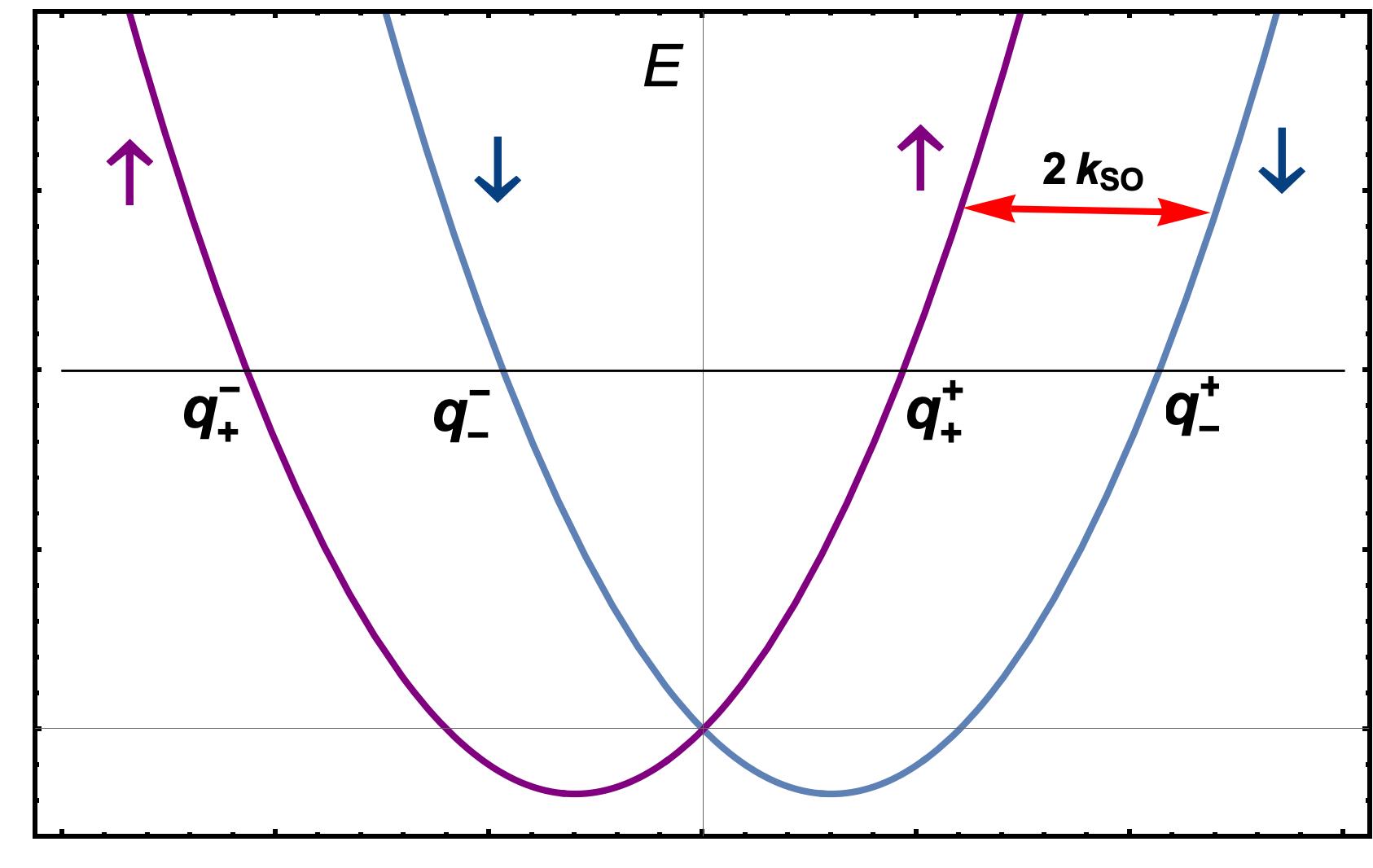}
    \caption{Dispersion relation for the one-dimensional spin-orbit coupled Hamiltonian. The meaning of the wave vectors $q_{\sigma^{\lambda}}$ is depicted. The momentum shifts between the dispersion is shown.}
    \label{RashbaDispersion}
\end{figure}

We take the incident beam to have an amplitude $\psi=\begin{pmatrix}\frac{1+s}{2}\\\frac{1-s}{2}\end{pmatrix}$,
where for $s=1$ corresponds to the up-spin normalized eigenstate of the $\sigma_z$ matrix and $s=-1$ to the down-spin state. While $s=0$ corresponds to the eigenstate in the $x$ direction. The normalization also allows access to all spin states in the $x-z$ plane of the Bloch sphere. 

We can faithfully rewrite the Hamiltonian in the following form
\begin{equation}
    {\cal H}=\frac{1}{2m}\left({\hat p}_x\mathds{1}_{\sigma}+m\Lambda(x)\sigma_y\right)^2+V_0-\frac{m\Lambda^2}{2},
\end{equation}
where we identify ${\cal A}_x=A_x^y\sigma_y=m\Lambda(x)\sigma_y$.
The velocity operator defined by $v_x=\partial {\cal H}/\partial p_x=((p_x/m)\mathds{1}_{\sigma}+\Lambda\sigma_y)$,
where we assume a single effective mass for the different scattering regions. Solving for the eigenvalues of this Hamiltonian we arrive at $E=\frac{1}{2m}\left(p_x+m \sigma\Lambda\right)^2-\frac{m\Lambda^2}{2}+V_0,$
where $\sigma=\pm 1$ is the spin quantum number. Equating $E=\hbar^2k^2/2m$ we define the wavevector outside the barrier region as $k$. Starting from the eigenvalue, we can solve for the possible values of $p_x=\hbar q$. A new quantum number arises that distinguishes right and left propagating waves. The resulting possible values of the wavevector under the barrier are
\begin{equation}
    q_{\sigma}^{\lambda}=\lambda\sqrt{k^2+k^2_{\rm so}-k^2_0}-\sigma k_{\rm so},
    \label{SOwavevector}
\end{equation}
and define $k_{\rm so}=m\Lambda/\hbar$ and $k_0^2=2m V_0/\hbar^2$. The meaning of the quantum numbers corresponding to Eq.\ref{SOwavevector} is depicted in Fig.\ref{RashbaDispersion}, where the degeneracy of two Kramer's pairs is evident. Note that for each direction of propagation, there are two distinct wavevectors with opposite spin labels and that the previous wave vector can be real or {\it complex} depending on the values of the incoming wavevector (with energy $E=\hbar^2k^2/2m$) and the height of the potential barrier.

\begin{figure}[ht]
\includegraphics[width=8.5cm]{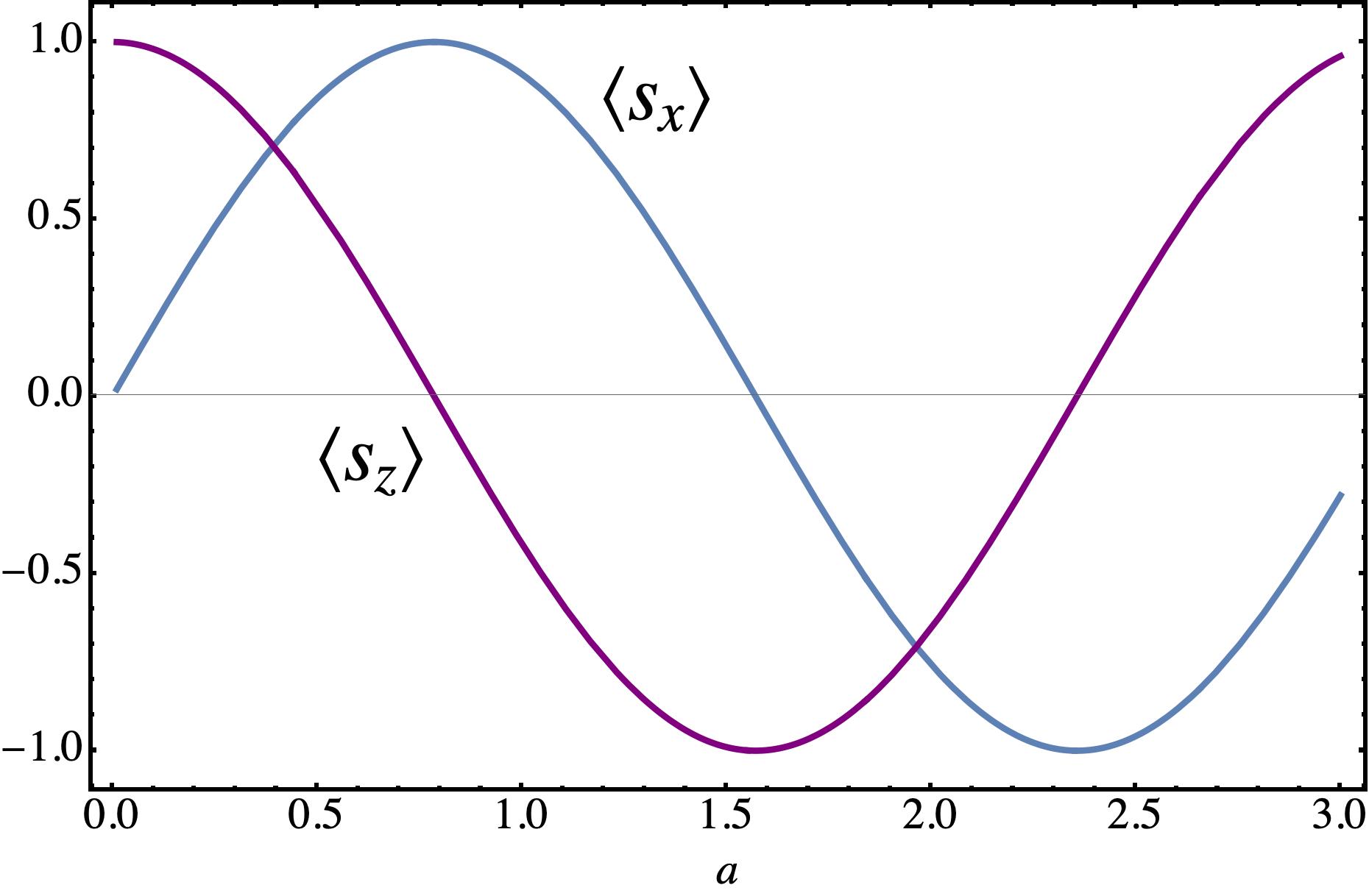}
    \caption{Precession of impinging spin in the $\hat z$ axis around $B_{SO}$. The k-vector under the barrier always has a real part, so the spin processes in the $x-z$ plane without relaxing toward the field.}
    \label{CoherentPrecession}
\end{figure}
As is easily derived from Eq.\ref{Hamiltonian2}, the Hamiltonian commutes with $\sigma_y$, and $\hat p_x$ so it has common eigenstates with $\sigma_y$ and the $\hat p_x$ eigenstates. In the $\sigma_z$ basis, the wavefunctions in the different regions are parameterized as follows
\begin{widetext}
\begin{eqnarray}
    \psi_1 &=&\begin{pmatrix}\frac{1+s}{2}\\\frac{1-s}{2}\end{pmatrix}e^{ikx}+\begin{pmatrix}A_+\\A_-\end{pmatrix}e^{-ikx},\nonumber\\
    \psi_2 &=& \frac{\alpha}{\sqrt{2}}\begin{pmatrix}1 \\i\end{pmatrix}e^{iq_+^{+}x}+\frac{\beta}{\sqrt{2}}\begin{pmatrix}1\\-i\end{pmatrix}e^{i q_-^{+}x}+\frac{\gamma}{\sqrt{2}} \begin{pmatrix}1\\i\end{pmatrix}e^{iq_+^{-}x}+\frac{\delta}{\sqrt{2}}\begin{pmatrix}1\\-i\end{pmatrix}e^{iq_-^{-}x},\\
    \psi_3&=&\begin{pmatrix}D_+\\D_-\end{pmatrix}\textcolor{black}{e^{ikx}},
\end{eqnarray}
\end{widetext}
where the appropriate $q^\lambda_\sigma$ wavevectors have implemented the coupling between the direction of propagation and spin orientation. The boundary conditions at the barrier limit $x_b$ are
\begin{eqnarray}
\psi_i(x_b)&=&\psi_{i+1}(x_b),\nonumber\\
{\hat v}_x\psi_i(x_b)&=&{\hat v}_x\psi_{i+1}(x_b),
\end{eqnarray}
where $i=1,2$, the index of the regions, and ${\hat v}_x=\left({\hat p}_x+m\Lambda\sigma_y\right)/m$. The second condition guarantees the continuity of the probability flux and not just the derivative of the wavefunction\cite{Molenkamp}.
The linear system of eight unknowns can be explicitly solved for the transmission and reflection probabilities.

The transmission of up-spin as a function of the entry spin polarization is
\begin{eqnarray}
    |t_+|^2&=&\frac{2 \Delta ^2 k^2 ((s-1) \cos (k_{so} a)-(s+1) \sin (k_{so}a))^2}{\Delta ^4+k^4+6 \Delta ^2 k^2-\left(k^2-\Delta ^2\right)^2 \cos (2 a \Delta )},\nonumber\\
    |t_-|^2&=&\frac{2 \Delta ^2 k^2 ((s-1) \sin (k_{so} a)+(s+1) \cos (k_{so} a))^2}{\Delta ^4+k^4+6 \Delta ^2 k^2-\left(k^2-\Delta ^2\right)^2 \cos (2 a \Delta )}\nonumber\\
\end{eqnarray}
where $\Delta=\sqrt{k^2+k_{so}^2-k_{0}^2}$. From here
we can see that $|t_{{\rm total}}|^2=|t_+|^2+|t_-|^2$ so the total conductance is
\begin{eqnarray}
    G_{{\rm total}}&=&G_++G_-=\frac{e^2}{h}|t_{{\rm total}}|^2\nonumber\\
    &=&\frac{8 e^2 \Delta ^2 k^2}{\Delta ^4+k^4+6 \Delta ^2 k^2-\left(k^2-\Delta ^2\right)^2 \cos (2 a \Delta )},\nonumber\\
    \end{eqnarray}
where $G_{\pm}=(e^2/h) |t_{\pm}|^2$ and $s^2=1$, which is spin independent\cite{Molenkamp,AharonyOraEntinComment} even in the presence of a barrier and for an open system.  Another way to appreciate this result is to see the incident up spin-oriented input state as an equal superposition of $|y_+\rangle$ and $|y_-\rangle$ states $(1 ~0)=1/\sqrt{2}(1 ~i)+1/\sqrt{2}(1~-i)$, eigenstates under the barrier. Computing then
\begin{equation}
P_y=\frac{|t_{+y}|^2-|t_{-y}|^2}{|t_{+y}|^2+|t_{-y}|^2}=\frac{2 \Im t^*_+ t_-}{|t_+|^2+|t_-|^2}=\langle s_y\rangle=0.
\label{PolarizationTransmission}
\end{equation}
The latter relations link the experimental and numerical formulas used in the literature for the polarization in terms of the expectation value of the spin\cite{ScipostTunneling}.

Fig.\ref{CoherentPrecession} shows how the average spin behaves on traversing the spin-active barrier region with SO coupling. In analogy with Buttiker's $U(1)$ magnetic field, we can define a new momentum-dependent magnetic field $B_{\rm SO}$
given the mapping $\lambda p_x\sigma_y=-\gamma {\bf B}_{\rm SO}\cdot {\bm \sigma}$ that results in ${\bf B}_{\rm SO}=-(\Lambda/\gamma) p_x {\bf u}_y $. $B_{\rm SO}$ lies in the negative $y$ direction for the model Hamiltonian. 
Precession follows correctly the torque equation $d\langle {\bf s\rangle}/dt=\gamma\langle{\bf s}\rangle\times {\bf B}_{\rm SO}$. Note that, as even under the barrier, the wavevector is complex, unlike the magnetic field case, precession proceeds with no generation of a spin component along the ${\bf B}_{\rm SO}$ direction. Also, both spin components suffer the same decay within the barrier (although dependent on SO) independent of their spin orientation (see Eq.[\ref{SOwavevector}]).  

The equal treatment of both spin projections renders a null polarization. We can also see the "spin-orbit magnetic field" $B_{SO}$ does not perform the same role as the $U(1)$ magnetic field under the barrier since up and down spins components do not have different decay rates. 

\section{Transmission with decoherence probe}

The spin-orbit coupling treats spin projections equally, so it cannot account for polarized spin polarization as expected in the CISS effect alone.  Nevertheless, the conditions for reciprocity are {\it not met} if, beyond the two terminal setups, there is a coupling to a third probe. A thermalization of electron transport to the environment through the electron-phonon or electron-electron interactions is inevitable at room temperature, providing time-reversal symmetry breaking and almost completing the analogy to a $U(1)$ magnetic field\cite{ScipostTunneling,ButtikerProbe}. The environment can be modeled as a lumped probe that disrupts the delicate coherences that yield the reciprocity theorem\cite{Kiselev,Bardarson2008} in the linear regime. This turns our attention to a tunneling molecular system to a three-probe scenario.

The Buttiker's voltage probe\cite{ButtikerProbe} is an ingenious way to introduce decoherence processes through the scattering matrix for an exactly solved model. It was designed for spinless electrons and was first used to introduce decoherence in free electron rings\cite{ButtikerProbe}, and generalized for spinors in reference\cite{Ellner}. 

In this section, we present a variation of the previously employed probe utilized in the context of persistent currents (as described in references \cite{Ellner, TorresPastawski, Damato}), with specific adaptations for the tunneling regime. It is worth noting that the probe we introduce is insensitive to spin, thereby avoiding the introduction of any additional spin-based selection mechanisms. To achieve this insensitivity, we incorporate two probes, each designed for a different spin species, both placed at the same location and connected to a third reservoir maintained at thermal equilibrium with a Fermi distribution at temperature $T$. Our probe is characterized as wide-band, supporting the wavevectors injected by the barrier channels.

The probe's scattering matrix is designed to generate an amplitude closely resembling a standard electron reservoir. Dealing separately with the spin components at any particular site under the barrier makes this coupling only approximately unitary but better as the barrier is higher than the input energy. This design approach disrupts interferences of the local fluxes corresponding to each spin orientation. As a direct consequence, introducing this probe breaks time-reversal symmetry (TRS), giving rise to equivalent effects of a magnetic field, which induces a net spin polarization in the direction of $B_{\rm SO}$.
\begin{figure}
    \centering
    \includegraphics[width=8.5cm]{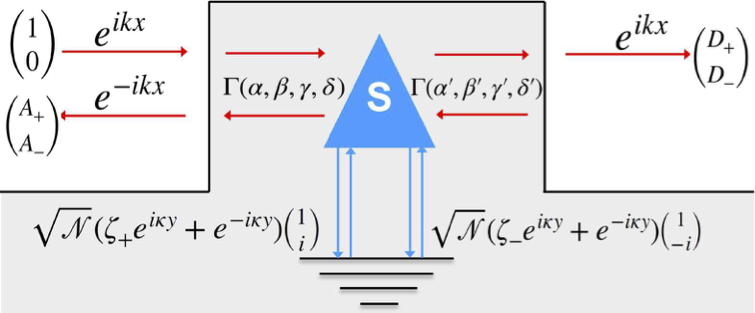}
    \caption{Buttiker's probe under the spin-orbit active barrier. The probe absorbs each eigenstate spin species under the barrier with the same scattering matrix, so no spurious spin selection is induced. Flux conditions are imposed on building the S matrix for a wideband Buttiker probe. Figure from reference\cite{ScipostTunneling}}
    \label{ButtikerProbe}
\end{figure}

The behavior of the Buttiker probe follows the combination of an ideal lead with $v=\hbar\kappa/m$ that supports a current $dI=e v(dN/dE)f(E)dE$ in the energy interval $dE$, where $f(E)$ is the Fermi distribution, $dN/dE=1/2\pi\hbar v$ is the density of states. This model can then induce level broadening\cite{Ellner,TorresPastawski}, (\cite{Damato} for Hamiltonian version) and level shifts under the barrier, and also depends on where the decoherence event occurs. Besides the coupling of the probe to the barrier, we can also control the temperature through the Fermi distribution of the attached reservoir. Our procedure for generating the $S$ matrix for each spin species is to couple the spin fluxes through the regular Buttiker matrix. Figure \ref{ButtikerProbe} shows the four regions that must be matched for continuity and flux. Under the barrier, the matching occurs at position $(x_0,y_0)=(x_0,0)$ where $y$ describes the coordinate of the third probe. The Scattering ($S$) matrix can then emulate a generic dephasing process\cite{Ellner}. 
For the case of the down spin coupling
\begin{widetext}
\begin{equation}
 \begin{pmatrix}
    \sqrt{\cal N}\zeta_-\kappa\\
    \Delta\beta'e^{iq_-^+x_0}\\
    -\Delta\delta e^{i q_-^-x_0}
    \end{pmatrix}=\begin{pmatrix}
    -(\mathcal{A}+\mathcal{B}) & \sqrt{\varepsilon} & \sqrt{\varepsilon} \\
     \sqrt{\varepsilon}& \mathcal{A} & \mathcal{B}\\
    \sqrt{\varepsilon} & \mathcal{B} & \mathcal{A}
    \end{pmatrix}
    \begin{pmatrix}
    \sqrt{\cal N}\kappa\\
    -\Delta\delta'e^{i q_-^-x_0}\\
    \Delta\beta e^{iq_-^+}
    \end{pmatrix}.
\end{equation}
\end{widetext}
$\mathcal{A}=(\sqrt{1-2\varepsilon}-1)/2$ and $\mathcal{B}=(\sqrt{1-2\varepsilon}+1)/2$, while ${{\cal N}=e f(E)dE/2\pi\hbar v}$ with $f(E)$ the Fermi distribution, $e$ the electron charge, $E$ the energy and $v$ the velocity of the carriers in the lead\cite{ButtikerProbe}. The parameter $\varepsilon$ takes the range $0<\varepsilon< 0.5$ and describes the coupling of the probe to the barrier from uncoupled to fully coupled. The labels follow the usage previously introduced where $q_{\sigma}^{\lambda}$, with $\sigma$ the spin label and $\lambda$ the sense of propagation label. Finally, $\kappa$ is the wavevector in the probe, and we have matched the fluxes between the barrier and the electron reservoir for the down spin component. 

We then rewrite the coupling Buttiker scattering matrix so that we are left with the vectors only in terms of the amplitudes $\beta',\delta,\beta,\delta'$. Note that the matching depends on the probe's position since the amplitude under the barrier depends on the coordinate $x_0$. We set $\kappa=\Delta$ since we assume a wideband probe, and we assume the injected wavevector from the barrier equals the wavevector in the reservoir-connected probe. We then obtain
\begin{widetext}
\begin{equation}
    \Psi_{out}=\begin{pmatrix}
    \sqrt{\cal N}\zeta_-\\
    \beta'\\
    \delta
    \end{pmatrix}=S_- \Psi_{in}=\begin{pmatrix}
    -(\mathcal{A}+\mathcal{B}) & -\sqrt{\varepsilon} e^{i q_-^-x_0} & \sqrt{\varepsilon} e^{i q_-^+x_0} \\
     \sqrt{\varepsilon} e^{-i q_-^+x_0}& -\mathcal{A} e^{-2i\Delta x_0} & \mathcal{B}\\
    -\sqrt{\varepsilon}e^{-iq_-^- x_0} & \mathcal{B} & \mathcal{A} e^{2i\Delta x_0}
    \end{pmatrix}
    \begin{pmatrix}
    \sqrt{\cal N}\\
    \delta'\\
    \beta
    \end{pmatrix},
    \label{ButtikerSMatrix}
\end{equation}
and for the up-spin
\begin{equation}
    \begin{pmatrix}
    \sqrt{\cal N}\zeta_+\\
    \alpha'\\
   \gamma
   \end{pmatrix}=\begin{pmatrix}
    -(a+b) & -\sqrt{\varepsilon} e^{i q_+^-x_0} & \sqrt{\varepsilon} e^{i q_+^+x_0} \\
    \sqrt{\varepsilon} e^{-i q_+^+x_0}& -\mathcal{A} e^{-2i\Delta x_0} & \mathcal{B}\\
   -\sqrt{\varepsilon}e^{-iq_+^- x_0} & \mathcal{B} & -\mathcal{A} e^{2i\Delta x_0}
   \end{pmatrix}
   \begin{pmatrix}
   \sqrt{\cal N}\\
   \gamma'\\
    \alpha
   \end{pmatrix},
\end{equation}
\end{widetext}
where $\Psi_{in,out}$ represents the input/output amplitudes to the junction, and $S_{\pm}$ is the scattering matrix for the up/down spin label.  
\begin{figure}
    \centering
    \includegraphics[width=8.5cm]{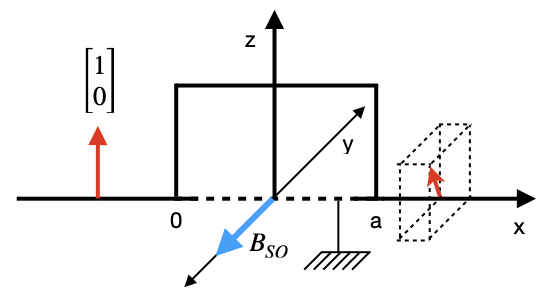}
    \caption{Spin tunneling setup chosen: We inject a spin-up electron in the $z$-axis into a spin-active barrier with a $B_{SO}$ in the $-{\hat y}$ direction. Asymmetric tunneling in the $y$ quantization axis produces a spin polarization or a net spin component in the $y$ direction.}
    \label{ButtikerPrecession}
\end{figure}


\begin{figure}
    \centering
    \includegraphics[width=8.0cm]{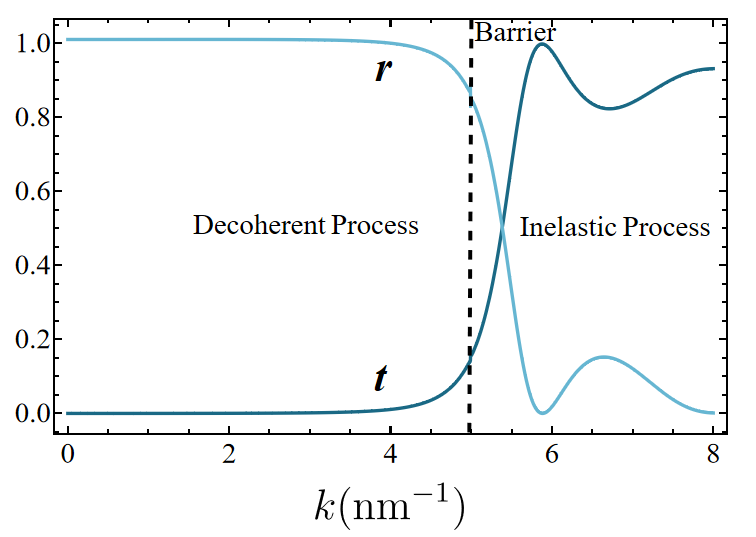}
    \includegraphics[width=7.7cm]{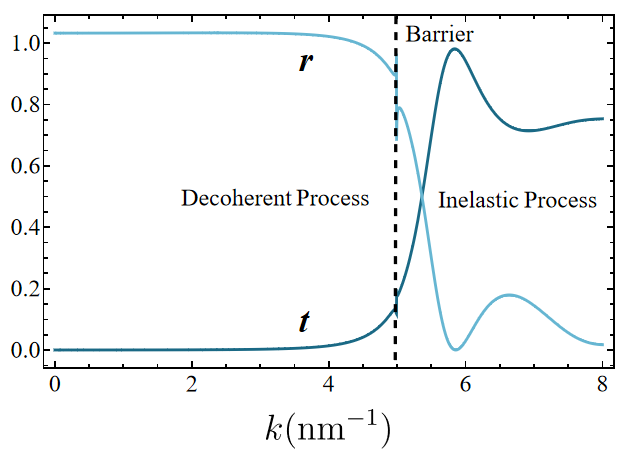}
    \caption{Top Panel: Transmission and reflection as a function of the input wavevector $k$. Unitarity above and below the barrier with our scattering matrix model based on the Buttiker probe. Here $\epsilon=0.005$, $k=1-8~nm^{-1}$, $k_{SO}=~nm^{-1}$, $k_0=5 nm^{-1}$, $a=1~nm$ in units of nm, $s=1$, $x_o=0.5~nm$ while $\mathcal{N} = 0.1$. Bottom panel: Transmission and reflection as a function of wavevector $k$ with barrier height set at $k_0=5~nm^{-1}$ but the coupling $\epsilon=0.05$. It can be seen that while below the barrier, the process remains almost unitary, there is a more pronounced non-unitarity/current leakage given the higher transmission amplitudes as compared to the tunneling process, 
    }
    \label{UnitaryConservation}
\end{figure}
\begin{figure}
    \centering
    \includegraphics[width=8.5cm]{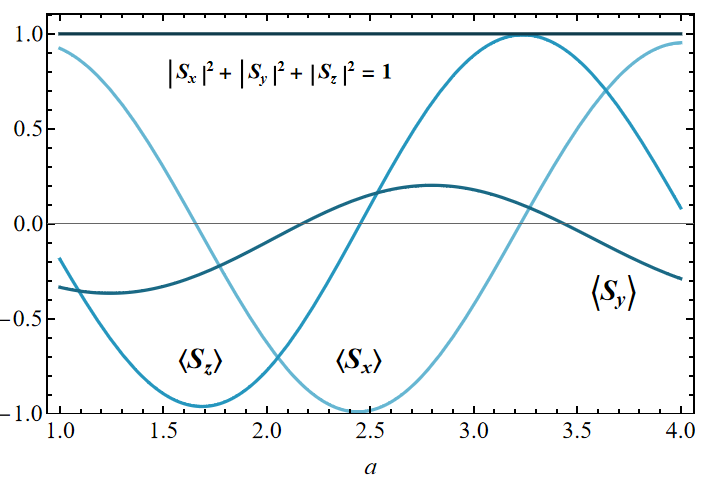}
    \caption{Spin precession when the decoherence probe couples to a particular point $x_0$ under the barrier. A disruption of spin precession in the $x-z$ plane is observed, generating a spin polarization analogous to an actual magnetic field\cite{ScipostTunneling} but with a relaxation direction depending on coupling strength and probe position. The norm of the spinor is preserved. The parameters depicted are $k=2~nm^{-1}$, $k_{so}=1~nm^{-1}$, $k_0=4~nm^{-1}$, $a=1-4~nm$, $s=1$, $x_0=0.8~nm$, $\epsilon=0.02$, $\mathcal{N} = 0.01$.}
    \label{ButtikerPrecessionDecoherence}
\end{figure}
\begin{figure}
    \centering
    \includegraphics[width=8.5cm]{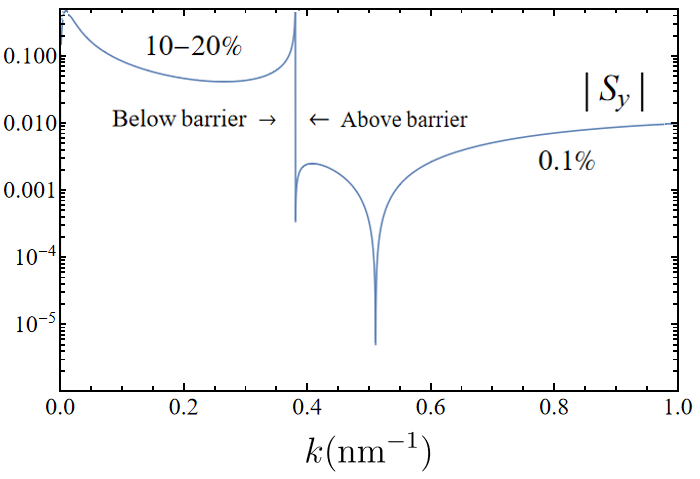}
    \caption{Spin Polarization as a function of the input wavevector $k$ with barrier height set at ($k_0=0.4nm^{-1}$) showing the difference in polarization of up to 3 orders for the polarization magnitude for the tunneling case versus polarization above the barrier.($k=0-1~nm^{-1}, k_{so}=0.1~nm^{-1}, k_0=0.4~nm^{-1}, a=5~nm, s=1, x_0=1.5~nm, \epsilon=0.01, \mathcal{N} = 0.1$)}
    \label{PolarizationAboveBelowBarrier}
\end{figure}

\begin{figure}
    \centering
    \includegraphics[width=8.5cm]{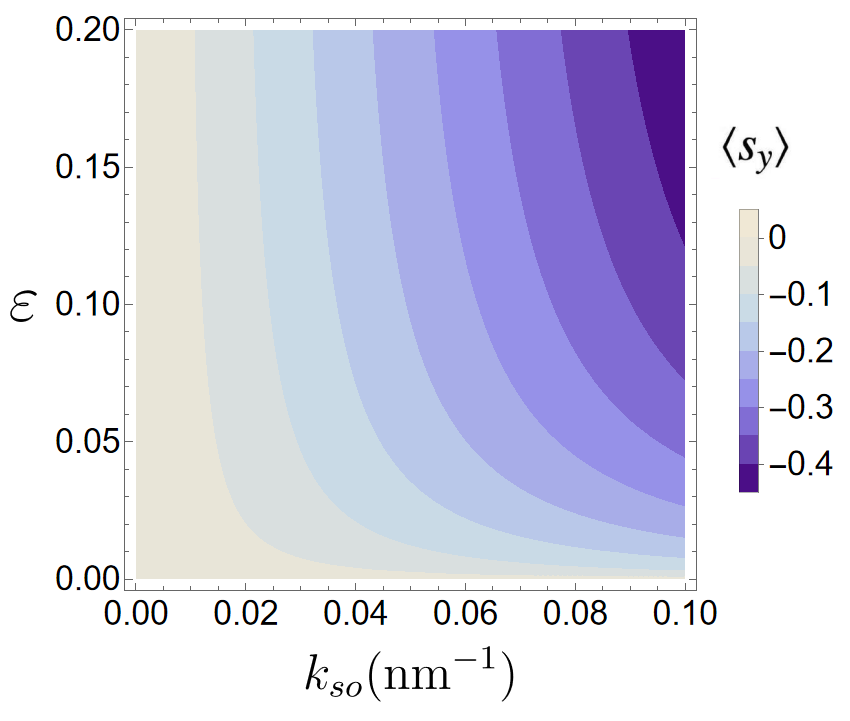}
    \includegraphics[width=8.5cm]{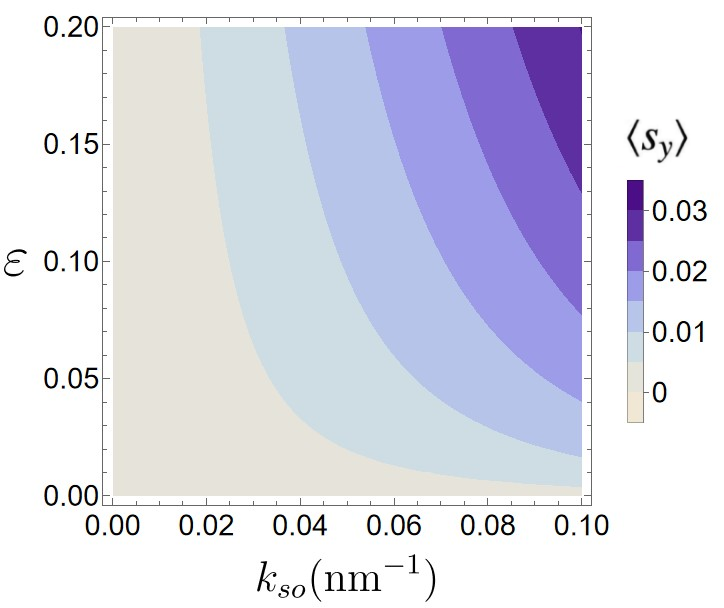}
    \caption{Top Panel: Spin polarization generated by decoherence analogous to that caused by a real external magnetic field. The contour plot shows the effect of a spin-orbit and decoherence coupling, consistent with estimates of ref.\cite{VarelaZambrano} for tunneling. The appearance of alignment of the spin to the $B_{\rm SO}$ is very sensitive to the coupling to the Buttiker probe. The parameter for under barrier: ($k=0.4~nm^{-1}, k_0=1~nm^{-1}, a=5~nm, s=1, x_0=1.5~nm, \mathcal{N} = 0.01$). Bottom Panel: For transmission above the barrier, the polarization is an order of magnitude weaker. The parameters here are ($k=0.4~nm^{-1}$, $k_0=0~nm^{-1}$, $a=5~nm$, $s=1$, $x_0=1.5~nm$, $\mathcal{N} = 0.01$).}
    \label{ContourEpsilonKsoAboveBelowBarrier}
\end{figure}
The spin scattering setup is depicted in Fig.\ref{ButtikerPrecession}, as in the coherent tunneling scenario, we inject spin up (in the $z$ quantization axis), which is decomposed into equal components in the basis functions of the SO term. Beyond the boundary conditions discussed at $x=0$ and $x=a$, we must include the matching equations \ref{ButtikerSMatrix} at point $x=x_0$. We can solve for this system exactly\cite{ScipostTunneling}. Fig.\ref{ButtikerPrecession} shows qualitatively that precession, only in the $x-z$ plane, amounts to equal treatment of both $y$ and $-y$ projections according to Eq.[\ref{PolarizationTransmission}]. This symmetry is broken here, yielding a ``relaxation" toward the $y$ axis, as depicted. This is analogous to what a magnetic field would achieve\cite{Buttiker,ScipostTunneling}. Figure \ref{UnitaryConservation} shows how the norm is preserved for a scattering process as we raise the energy from below to above the barrier. Below the barrier, our matching model to the Buttiker-probe works very well, preserving the norm, while above the barrier, it yields probability leakage,  especially at the larger coupling. Thus, the zero current condition is not strictly satisfied. Pending an exact formulation of the probe with spin, we draw conclusions for this approximate model of pure dephasing.

Figure \ref{ButtikerPrecessionDecoherence} shows the appearance of the $y$ component of spin as a function of the barrier length in the presence of the decoherence probe. As in the coherent case, we show an extensive range of parameters to illustrate how precession is connected to polarization. Note also that polarization does not "relax" toward the particular $B_{SO}$ direction ($-y$) as in a magnetic field, as can be seen in the figure. The specific orientation on the $B_{SO}$ direction depends on the details of interference effects introduced by the probe\cite{ScipostTunneling}.

\begin{figure}
    \centering
    \includegraphics[width=8.5cm]{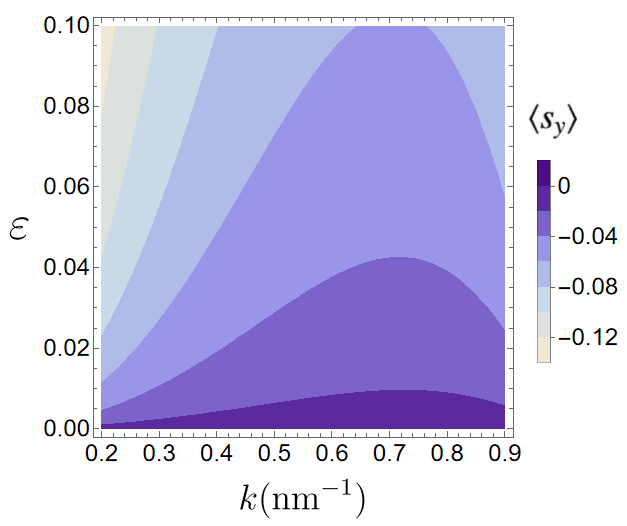}
    \includegraphics[width=8.5cm]{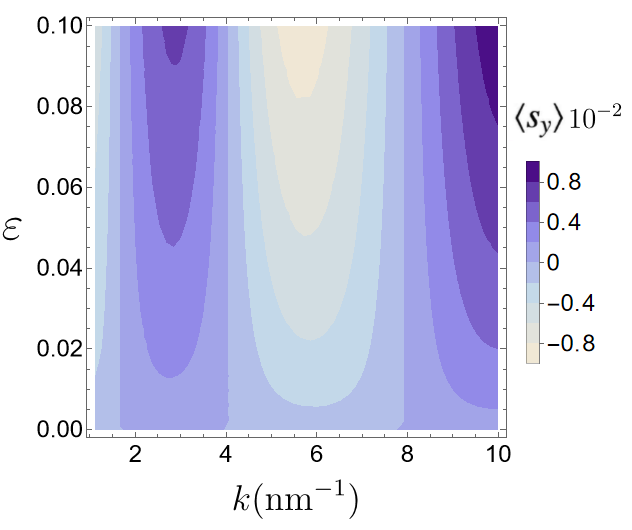}
    \caption{Top Panel: Spin polarization generated by decoherence under the barrier. Polarization is stronger and only antialigned to the field due to the tunneling effect. Bottom Panel: Spin polarization above the barrier. Polarization is weaker and oscillates between alignment and anti-alignment to the $B_{\rm SO}$ field. Here ($k_{so}=0.04~nm^{-1}, k_0=1~nm^{-1}, a=4~nm, s=1, x_o=0.8~nm, \mathcal{N} = 0.1$)}
    \label{ContourEnergyDependence}
\end{figure}

As the role of the spin active barrier is not enough to produce relaxation, as happens in the presence of a magnetic field, it is possible to achieve time-reversal symmetry breaking by coupling the Buttiker probe with energies above or below the barrier and in the absence of the barrier. Figure \ref{PolarizationAboveBelowBarrier} shows the magnitude of $s_y$ as one increases the input energy from below to above the barrier. While below the barrier, the polarization reaches between 10-20\%, above the barrier, the polarization is much lower by two orders of magnitude. The peaks in the figure correspond to the effects of precession and the energy passing the barrier height.  Figure \ref{ContourEpsilonKsoAboveBelowBarrier} compares the polarization power of scattering with barrier and without barrier as a function of the coupling to the Buttiker probe and the strength of the SO coupling. Finally, Figure\ref{ContourEnergyDependence} shows the polarization strength for scattering above and below the barrier as a function of the coupling strength to the Buttiker probe and the input energy. As previously discussed, one can see that while both situations polarize spin below the barrier polarizes orders of magnitude higher, both cases can polarize spin as an interference effect that can change direction depending on the details of the input energy and coupling to the third probe.

\section{Discussion and Summary}

In the preceding sections, we have worked out a one-dimensional scattering off a barrier featuring intrinsic spin-orbit (SO) interaction. This model can be derived from a three-dimensional system by disregarding the orbital degree of freedom\cite{VarelaZambrano}. The origin of spin-orbit coupling lies in the spatial configuration of the $p$ orbitals within the helical structure of the model. Without the latter configuration, the strength of the spin-orbit coupling would be significantly diminished by orders of magnitude\cite{Varela2016}, much like the comparison between spin-orbit coupling in planar graphene and carbon nanotubes as explored in reference\cite{Huertas2006}.

The spin-orbit coupling in this model is of atomic origin, while the spin and orbital degrees of freedom within the helix remain uncoupled. Consequently, the orbital degree of freedom only influences the kinetic energy. It introduces orbital angular momentum, which, as described in the reference \cite{LopezVarelaMedina}, can enhance the effects of spin-orbit coupling (see also a possible geometrical SO coupling\cite{GeometricalSO}). Thus, chirality plays a crucial role in our model of CISS, contributing to its approximately 1-10 meV strength. In a two-terminal setup at zero temperature, the chiral structure and spin-orbit interaction alone do not suffice for spin selectivity\cite{Kiselev,Bardarson2008,Bart1}. As reviewed\cite{ScipostTunneling} here, the spin-orbit coupling only induces spin precession due to the spin torque produced by the spin-orbit ``magnetic field" ($B_{SO}$), without any asymmetric treatment of distinct spin species. Here, we have shown the equivalence of the spin asymmetry definition used in the literature and the appearance of a spin component perpendicular to the precession plane called the spin polarization. 

One of the most remarkable facets of the CISS effect is that it is at room temperature effect, thus its possible relevance to biological systems. Fully coherent quantum models offer only a partial explanation, prompting the introduction of various hypotheses to account for temperature-related effects. These hypotheses include interactions with equilibrated populations of phonon or electron baths or lossy, non-unitary systems. Here, we model such effects by generalizing a Buttiker probe with two sources of decoherence: i) Injection of incoherent phases in the spin-orbit active range and treating the spin components incoherently with no bias for either component. The model works as a very good purely dephasing probe under the barrier, but the model is lossy or inelastic above the barrier or with no barrier.  

Tunneling is a typical electron transfer mechanism in large molecules through, for example, polaron motion\cite{Michaeli}. We report here that tunneling plus decoherence is more effective in generating spin polarization than otherwise. This finding is significant since most models assume free electron transport on helical structures. We thus report very high polarization for reasonable polaron-type tunneling with physically reasonable parameters.

While trying to build in the ingredients for strong spin polarization, we have built a model that selects a quantization axis that corresponds to the direction of the spin-orbit ``magnetic field", but that does not break time-reversal symmetry, so decoherence is introduced as a generic TRS breaking mechanism. Yet the latter two ingredients still do not render a proper magnetic field since there is only relaxation onto the quantization axis, with an alignment that depends on the interference detail produced by the Buttiker probe. This was illustrated by our results, which show that changing the input energy under the barrier allows one to change the spin alignment onto the $B_{SO}$ quantization axis. Although the stronger polarization for the tunneling case can make a change of alignment more difficult to observe for reasonable parameters, the alignment is nevertheless non-monotonous, which indicates an interference effect.

In more realistic models, coupling the two-terminal model with electron-phonon/electron interactions could play the decoherence role, yielding different smoking gun temperature dependencies and shedding light on the mechanism responsible for breaking time-reversal symmetry. As a result, thermal effects provide a coherent framework for fully understanding the CISS effect. In the mechanism we have outlined here, the disruption of spin precession directly introduces an asymmetry in the spatial dependence of spin amplitudes, thus determining spin selection. This mechanism results in changing polarizations linked to the thermalization process.







\section*{Acknowledgments}
M.M, acknowledges the hospitality of The Chair in Material Science and Nanotechnology, TU Dresden, where part of this work was done. E.M. Acknowledges support from Poligrant 17945 USFQ and an invited professorship at Lorraine University, where this work began. B.B. Acknowledges support from LA-CoNGA physics (Latin American Alliance for Capacity buildiNG in Advanced Physics). 
%


\clearpage



\end{document}